# Errorless Codes for Over-loaded CDMA with Active User Detection

Pedram Pad, Mahdi Soltanolkotabi, Saeed Hadikhanlou, Arash Enayati and Farokh Marvasti

Advanced Communications Research Institute (ACRI),
Department of Electrical Engineering, Sharif University of Technology
Tehran, Iran
Email: {pedram_pad, msoltan, s_hadikhanlou}@ee.sharif.edu, arashenayati@gmail.com and marvasti@sharif.edu

***Abstract*—In this paper we introduce a new class of codes for over-loaded synchronous wireless CDMA systems which increases the number of users for a fixed number of chips without introducing any errors. In addition these codes support active user detection. We derive an upper bound on the number of users with a fixed spreading factor. Also we propose an ML decoder for a subclass of these codes that is computationally implementable. Although for our simulations we consider a scenario that is worse than what occurs in practice, simulation results indicate that this coding/decoding scheme is robust against additive noise. As an example, for 64 chips and 88 users we propose a coding/decoding scheme that can obtain an arbitrary small probability of error which is computationally feasible and can detect active users. Furthermore, we prove that for this to be possible the number of users cannot be beyond 230.**

## I. INTRODUCTION

In Direct-Sequence Code Division Multiple Access (DS-CDMA) due to practical conditions it is desirable to use binary antipodal signatures (spreading codes) in conjunction with BPSK modulation. In these systems we can obtain errorless transmission using orthogonal codes (e.g. Hadamard codes) under the assumption of noiseless channel. This is only true if the number of users does not exceed the spreading factor (under-loaded or fully-loaded CDMA). When the number of users exceeds the spreading factor, such orthogonal codes do not exist. Also, using Pseudo-Noise (PN) spreading signatures creates interference that cannot be completely removed and results in errors in the Multi-User Detection (MUD) receiver [1-3]. There are papers that discuss double orthogonal codes for increasing capacity [4-5]. These codes are non-binary complex codes (equivalent to $m$ phases for MC-OFDM) and thus are not a fair comparison to binary codes.

In bandwidth limited channels, over-loaded CDMA is required. Most of the research in the over-loaded case is focused on code design and Multi-Access Interference cancellation for decreasing the probability of error. Examples of these type of research are pseudo random spreading (PN) codes [6-7]; OCDMA/OCDMA (O/O) codes [8-10], Multiple-OCDMA (MO) codes [11], and Binary Welch Bound Equality (BWBE) codes [12-14]. None of the signatures and decoding schemes introduced above guarantee errorless communication in an ideal (noiseless) synchronous channel in the over-loaded case. All of these codes were designed with the idea of minimizing the criterion of Total Squared Correlation (TSC) in mind. According to [15] minimizing TSC maximizes channel capacity when the input distribution is Gaussian. However, when the input alphabet is $\pm 1$ these codes do not necessarily maximize the channel capacity and thus may not result in a lower probability of error.

In [16] we presented a new class of codes named COW. Contrary to the aforementioned codes these codes can achieve errorless transmission in the ideal (noiseless) channel. Although the COW codes can achieve high over-loading factors[1], we need to know the active users for proper decoding.

In many random access communication systems, identification of the active users will help increase the system capacity as noted in [17]. For example in ad-hoc networks as observed in [18], "Optimal transmission strategies require the identification and localization of active nodes in the neighborhood of the transmitter".

In this paper we present a new set of over-loaded codes which guarantee errorless transmission in noiseless channels and are also robust against additive noise. In addition they are capable of detecting active users. We call this new class of codes, Codes for Over-loaded Wireless CDMA with Detection of Active users (COWDA). These codes are well suited for synchronous Code Division Multiplexing (CDM) in broadcasting and downlink wireless applications. Due to their active user detection capability they can also be used in spatial multiplexing applications and ad-hoc networks. Furthermore, these codes can result in bandwidth saving by the use of lower chip rates. In addition, we will propose a decoding algorithm at the receiver that is ML and computationally feasible. As an example, for a signature length of 64 these new codes can achieve over-loading factors of about %38 that can be practically decoded in time. Furthermore, we have proven the existence of codes with an overload factor of almost %48. We will also obtain an upper bound on the over-loading factor, where it cannot be beyond %260.

In section II necessary and sufficient conditions for errorless transmission with active user detection in over-loaded CDMA systems are discussed. Also methods of constructing large COWDA codes with a high percentage of over-loading factors will be presented. In section III an upper bound on the number of users is obtained for a given spreading factor. The decoding algorithm is presented in section IV. Simulation results appear in section V.

[1] The percentage of the number of users divided by the number of chips minus 1.

## II. Codes for Over-loaded Wireless CDMA with Detection of Active users (COWDA)

For developing COWDA codes, we will first present an intuitive geometric interpretation and then develop the codes mathematically. At a given time the multi-user binary data can be represented by an $n$-dimensional vector (with $\pm 1$ denoting active user data and 0 denoting non-active users); these vectors can be interpreted as a set of discrete points on the vertices and inside an $n$-dimensional hyper-cube ($\{-1,0,+1\}^n$). The data pertaining to the active users are multiplied by their respective $m$-chip long binary antipodal signatures and finally their summation is transmitted. Thus, the transmitted $m$-tuple vector can be viewed as the result of the multiplication of an $m \times n$ matrix (with columns being the signatures of different users) by the input $n$-dimensional vector. Alternatively, this can be viewed as a linear mapping of the points on the vertices and inside the hyper-cube onto points in an $m$-dimensional space ($m < n$). As long as, the resulting points in the $m$-dimensional space are distinct, the mapping is one-to-one and therefore, we can uniquely decode each received $m$-tuple vector at the receiver. However, if the points in the $m$-dimensional space are not distinct, the mapping is not one-to-one and thus not invertible. This results in irremovable interference. Consequently, we look for codes that map the points in the $n$-dimensional space onto distinct points in the $m$-dimensional space. Most of the over-loaded codes discussed in the literature do not possess this property and thus their MUD scheme cannot be perfect. We call the class of codes with the above mentioned property Codes for Over-loaded Wireless CDMA with Detection of Active Users (COWDA). Now we will develop a systematic method for the generation of such codes.

**Lemma 1** Denote the $n$-dimensional input vectors $\{-1,0,+1\}^n$ with the set $\boldsymbol{V}$. The necessary and sufficient condition for the multiplication of a $\{\pm 1\}$-matrix $\boldsymbol{C}$ with elements of $\boldsymbol{V}$ to be a one-to-one mapping is that $Ker\boldsymbol{C} \cap \{0,\pm 1,\pm 2\}^n = \{0\}^n$, where $Ker\boldsymbol{C}$ is the null space of $\boldsymbol{C}$.
*The proof is trivial.*

**Corollary 1** If $\boldsymbol{C}$ is a COWDA matrix then:
a- New COWDA matrices can be generated by multiplying each row or column of $\boldsymbol{C}$ by $-1$.
b- New COWDA matrices can be generated by permutations of columns or rows of $\boldsymbol{C}$.
c- By adding an arbitrary binary antipodal row to $\boldsymbol{C}$, we obtain another COWDA matrix.
*The proof is trivial.*

**Note 1** To verify that a $\{\pm 1\}$-matrix is COWDA using Lemma 1 it is sufficient to check $5^n$ vectors. Ignoring the zero-vector and considering the fact that half of the vectors are the negative of the other half, we need to search only amongst $(5^n - 1)/2$ vectors, which is a very huge number. Now we suggest a method that can decrease this number dramatically. Partition $\boldsymbol{C}$ as $[\boldsymbol{A}\ \boldsymbol{B}]$ where $\boldsymbol{A}$ is an $m \times m$ invertible matrix (according to Corollary 1 this can be done in most cases). Suppose there exists a vector $X \in \{0,\pm 1,\pm 2\}^n$ such that $\boldsymbol{C}X = 0$. Then there exists vectors $X_1$ and $X_2$ of size $m$ and $(n-m)$ respectively, with entries in $\{0,\pm 1,\pm 2\}$ such that $\boldsymbol{A}X_1 + \boldsymbol{B}X_2 = 0$. Consequently, we only need to search amongst $(5^{n-m} - 1)/2$ likely vectors $X_2$ in $\{0,\pm 1,\pm 2\}^{n-m}$ and check that $-\boldsymbol{A}^{-1}\boldsymbol{B}X_2$ belongs to $\{0,\pm 1,\pm 2\}^m$. For example, to check that the $\boldsymbol{C}_{16\times 22}$ matrix of Table. 1 is COWDA we only need to search among $(5^6 - 1)/2$ vectors.

**Theorem 1** Assume that $\boldsymbol{C}$ is an $m \times n$ COWDA matrix and $\boldsymbol{P}$ is an invertible $k \times k$ $\{\pm 1\}$-matrix, then $\boldsymbol{P} \otimes \boldsymbol{C}$ is a $km \times kn$ COWDA matrix, where $\otimes$ denotes the Kronecker matrix product.
*Proof*: Obviously, $\boldsymbol{P} \otimes \boldsymbol{C}$ is a $\{\pm 1\}$-matrix. Suppose that $(\boldsymbol{P} \otimes \boldsymbol{C})Z = 0$ where $Z$ has entries in $\{0,\pm 1,\pm 2\}$. Multiplying both sides of this equation by $(\boldsymbol{P}^{-1} \otimes \boldsymbol{I}_m)$, we have

$$(\boldsymbol{I}_k \otimes \boldsymbol{C})Z = 0$$

The above set of equations can be decoupled into $k$ different equations of the form

$$\boldsymbol{C}Z_i = 0$$

where $Z_i$ denotes the $\left((i-1)n+1\right)^{\text{th}}$ to the $(in)^{\text{th}}$ entries of $Z$ with $1 \le i \le k$. However, we know that $\boldsymbol{C}$ is COWDA, thus according to Lemma 1, $Z_i$ equals the zero vector for $1 \le i \le k$. Hence, $Z$ is the zero-vector and again according to Lemma 1 $\boldsymbol{P} \otimes \boldsymbol{C}$ is COWDA. ■

In the following theorem we will prove the existence of COWDA codes with a higher percentage of the over-loading factor.

**Theorem 2** Assume $\boldsymbol{C}$ is an $m \times n$ COWDA matrix and $\boldsymbol{H_4}$ is a $4 \times 4$ Hadamard matrix. We can add $\lceil (m-1)log_5^2 \rceil$ columns to $\boldsymbol{H_4} \otimes \boldsymbol{C}$ to obtain another COWDA matrix.
*This theorem is proved in the Appendix.*

**Note 2** $n/m \to \infty$ as $m \to \infty$.
This observation is a direct result of Theorem 2 since $n/m$ is of order $O(\log m)$. This implies that as the chip rate increases the number of users grow much faster.

$$\begin{bmatrix}
+&+&+&+&+&+&+&+&+&+&+&+&+&+&+&+&+&+&+&+&+&+\\
+&-&+&-&+&-&+&-&+&-&+&-&+&-&+&-&-&-&+&-&+&+\\
+&+&-&-&+&+&-&-&+&+&-&-&+&+&-&-&+&-&-&-&+&-\\
+&-&-&+&+&-&-&+&+&-&-&+&+&-&-&+&+&+&-&-&+&+\\
+&+&+&+&-&-&-&-&+&+&+&+&-&-&-&-&-&-&-&-&-&-\\
+&-&+&-&-&+&-&+&+&-&+&-&-&+&-&+&+&-&+&-&+&+\\
+&+&-&-&-&-&+&+&+&+&-&-&-&-&+&+&-&+&+&+&+&-\\
+&-&-&+&-&+&+&-&+&-&-&+&-&+&+&-&-&-&-&+&+&-\\
+&+&+&+&+&+&+&+&-&-&-&-&-&-&-&-&-&-&+&-&-&-\\
+&-&+&-&+&-&+&-&-&+&-&+&-&+&-&+&+&-&+&-&-&+\\
+&+&-&-&+&+&-&-&-&-&+&+&-&-&+&+&+&+&+&-&-&+\\
+&-&-&+&+&-&-&+&-&+&+&-&-&+&+&-&-&+&+&+&+&+\\
+&+&+&+&-&-&-&-&-&-&-&-&+&+&+&+&+&+&-&-&-&-\\
+&-&+&-&-&+&-&+&-&+&-&+&+&-&+&-&+&-&+&-&+&+\\
+&+&-&-&-&-&+&+&-&-&+&+&+&+&-&-&+&-&-&+&+&-\\
+&-&-&+&-&+&+&-&-&+&+&-&+&-&-&+&+&+&+&-&-&+
\end{bmatrix}$$

Table. 1. $\boldsymbol{C}_{16\times 22}$ where + denotes +1 and − denotes −1. Notice that the first 16 columns of the above matrix is a $16 \times 16$ Hadamard matrix.

**Example 1** In the first step, applying Theorem 2 on a $4 \times 4$ Hadamard matrix, we first get a $16 \times 18$ COWDA matrix ($\boldsymbol{C}_{16\times18}$). By computer search we also found a $16 \times 22$ COWDA matrix ($\boldsymbol{C}_{16\times22}$ which is shown in Table. 1). According to Theorem 1, $\boldsymbol{C}_{16\times22}$ leads to a $64 \times 88$ COWDA matrix by the Kronecker product $\boldsymbol{H}_4 \otimes \boldsymbol{C}_{16\times22}$ (where $\boldsymbol{H}_4$ is a $4 \times 4$ Hadamard matrix); this implies that we can have errorless decoding for 88 users with only 64 chips; which has an over-loading factor of about %38 (we will introduce a suitable decoder for this code in section IV). However, reuse of Theorem 2 for $\boldsymbol{C}_{16\times22}$ leads to a $64 \times 95$ COWDA matrix ($\boldsymbol{C}_{64\times95}$). This implies an over-loading factor of %48.

## III. Upper Bound for the Over-Loading Factor

The following theorem provides an upper-bound for the over-loading factor of a COWDA matrix.

**Theorem 3** If $\boldsymbol{C} = [c_{ij}]$ is a COWDA matrix with $n$ columns (users) and $m$ rows (chips), then

$$n \leq -m\left(\sum_{k=-n}^{n} \frac{f(n,k)}{3^n} log_3 \frac{f(n,k)}{3^n}\right)$$

where

$$f(n,k) = \sum_{r=0}^{\lfloor\frac{n-k}{2}\rfloor} \binom{n}{r}\binom{n-r}{r+k}.$$

*Proof*: Suppose $Y = \boldsymbol{C}X$, where $\boldsymbol{C}$ is a COWDA matrix and $X$ is a random variable from the set $\{-1,0,+1\}^n$ with uniform distribution. Since the code matrix ($\boldsymbol{C}$) denotes a one-to-one transformation between $X$ and $Y$, then the vectors $Y$ and $X$ have the same amount of information. Thus,

$$H(Y) = H(X) = n\ log_2 3$$

where $H$ denotes the entropy function in bits. Since the entries of $X$ are from the set $\{-1,0,+1\}$ and the entries of the code matrix are $\pm1$, the entries of $Y$ are integers between $-n$ and $n$. The probability that an entry in $Y$ takes the value $k$ is equal to,

$$P(y_i = k) = \frac{f(n,k)}{3^n}$$

where $f(n,k)$ is the number of solutions of the equation,

$$x_1 + x_2 + \ \dots + x_n = k$$

with $x_i$ belonging to the set $\{-1,0,+1\}$. $f(n,k)$ can be calculated from the following formula,

$$f(n,k) = \sum_{r=0}^{\lfloor\frac{n-k}{2}\rfloor} \binom{n}{r}\binom{n-r}{r+k}$$

each entry in the above summation is the number of such solutions with $r$ 1's (there are $\binom{n}{r}$ such entries) and $r+k$ entries of $-1$'s (there are $\binom{n-r}{r+k}$ such entries). Knowing the probability of the single entries of the vector $Y$ ($P(y_i = k)$), we can calculate an upper bound for $H(Y)$ with the assumption of independence between the entries of $Y$. Thus,

$$H(Y) \leq -m\left(\sum_{k=-n}^{n} \frac{f(n,k)}{3^n} log_2 \frac{f(n,k)}{3^n}\right)$$

Therefore,

$$n\ log_2 3 \leq -m\left(\sum_{k=-n}^{n} \frac{f(n,k)}{3^n} log_2 \frac{f(n,k)}{3^n}\right)$$

which implies the upper bound in the theorem.■

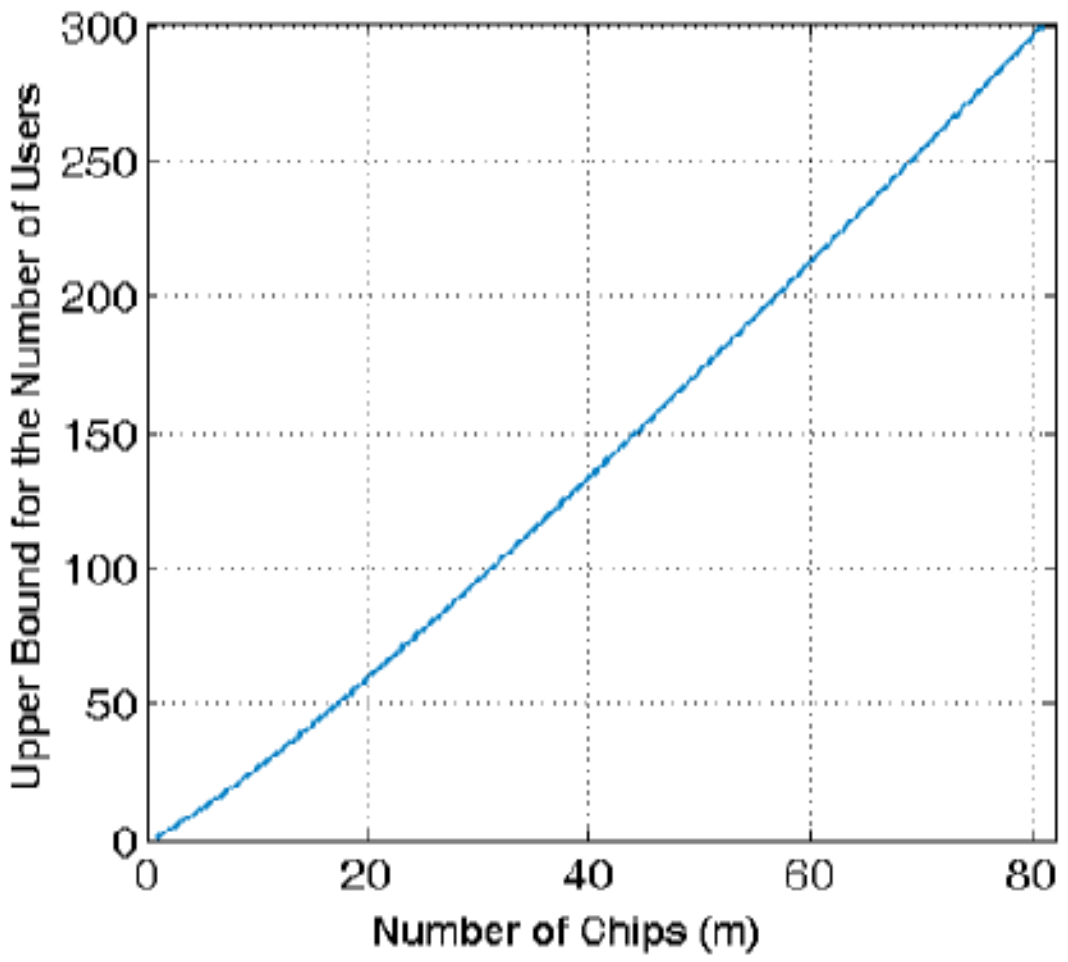

Fig.1. Upper bound on the number of users vs. the number of chips

The upper bound stated in Theorem 3 is shown in Fig. 1. This figure shows that we cannot have errorless communication with 64 chips and beyond 230 users, which implies an over-loading factor lower than %260.

## IV. Decoding Algorithm

In this section we will present a decoding algorithm for the proposed class of codes. This decoding scheme has the added advantage that it works with an unknown number of active users.

At the receiver a combination of the signatures of different users embedded in AWGN is received. This vector can be modeled as

$$Y = \boldsymbol{C}X + N$$

where $\boldsymbol{C}$ is the code matrix and $N$ denotes the noise vector which has a Gaussian distribution, with zero mean, and auto-covariance matrix $\sigma^2\boldsymbol{I}$ (where $\boldsymbol{I}$ denotes the identity matrix). To implement ML decoding for each user, it must minimize $\left\|Y - \boldsymbol{C}\hat{X}\right\|_2$, where the user's entry in $\hat{X}$ is $\pm1$ and the rest of the entries (of $\hat{X}$) belong to the set $\{-1,0,+1\}$. The reason behind this fact, is that each active user is not aware of the status of other users; it only knows that its own signature contributes to the received vector by a $\pm1$ occurrence and not 0. Consequently the user has to choose between $2 \times 3^{n-1}$ input vectors $\hat{X}$ (with $n$ denoting the number of users). The computational complexity of this implementation of the ML decoder is tremendously high. In the following we will present a decoding method with much lower complexity. We will also derive conditions under which this decoder is ML. This is accomplished in two major steps.

In the first step, we show that if the codes have been generated according to Theorem 1, the decoding problem can be reduced to a set of decoding problems with smaller code matrices. Consider a COWDA code matrix $\boldsymbol{C}_{rl\times rk} = \boldsymbol{P}_{r\times r} \otimes \boldsymbol{D}_{l\times k}$ generated by the Kronecker product of an invertible matrix $\boldsymbol{P}$ with a smaller COWDA matrix $\boldsymbol{D}$ (according to Theorem 1). The received vector is

$$Y = \boldsymbol{C}X + N = (\boldsymbol{P} \otimes \boldsymbol{D})X + N$$

Multiplying both sides by $(\boldsymbol{P}^{-1} \otimes \boldsymbol{I}_l)$ we have

$$(\boldsymbol{P}^{-1} \otimes \boldsymbol{I}_l)Y = (\boldsymbol{P}^{-1} \otimes \boldsymbol{I}_l)\big((\boldsymbol{P} \otimes \boldsymbol{D})X + N\big) = (\boldsymbol{I}_r \otimes \boldsymbol{D})X + (\boldsymbol{P}^{-1} \otimes \boldsymbol{I}_l)N$$

This implies that the first $l$ elements of $(\boldsymbol{P}^{-1} \otimes \boldsymbol{I}_l)Y$ depends only on the first $k$ elements of $X$ plus the new noise term $(\boldsymbol{P}^{-1} \otimes \boldsymbol{I}_l)N$, the second $l$ elements of $(\boldsymbol{P}^{-1} \otimes \boldsymbol{I}_l)Y$ depends only on the second $k$ elements of $X$ plus the new noise term $(\boldsymbol{P}^{-1} \otimes \boldsymbol{I}_l)N$, and so on. Hence, for extracting the first $k$ bits of $X$ it's sufficient to consider only the first $l$ elements of $(\boldsymbol{P}^{-1} \otimes \boldsymbol{I}_l)Y$, for extracting the second $k$ bits of $X$ it's sufficient to consider only the second $l$ elements of $(\boldsymbol{P}^{-1} \otimes \boldsymbol{I}_l)Y$, and so on. Thus, we have divided the problem of decoding a CDMA system with $m = rl$ chips and $n = rk$ users to decoding $r$ CDMA systems with $l$ chips and $k$ users. This results in a huge reduction of computational costs. If the matrix $\boldsymbol{P}$ is Hadamard, the ML decoder of the bigger system becomes equivalent to the ML decoders of the smaller systems (because $\sqrt{r}\boldsymbol{P}^{-1} \otimes \boldsymbol{I}_l$ is a unitary matrix and does not change the auto-covariance matrix of noise).

In the second major step, we will further reduce the complexity of the smaller systems. Consider $\boldsymbol{D} = [\boldsymbol{A} \quad \boldsymbol{B}]$ where $\boldsymbol{A}$ is an $l \times l$ invertible matrix and $\boldsymbol{B}$ is an $l \times (k - l)$ matrix. The reason that $\boldsymbol{A}$ can be considered invertible is that the assumption of $\boldsymbol{D}$ being full rank is not very restricting and due to Corollary 1, columns of $\boldsymbol{D}$ can be permuted. Using this partitioning

$$Y = \boldsymbol{D}X + N = [\boldsymbol{A} \quad \boldsymbol{B}]\begin{bmatrix} X_1 \\ X_2 \end{bmatrix} + N = \boldsymbol{A}X_1 + \boldsymbol{B}X_2 + N$$

where $X_1$ and $X_2$ are $l \times 1$ and $(k - l) \times 1$ vectors, respectively. Multiplying both sides by $\boldsymbol{A}^{-1}$, we arrive at the equation:

$$\boldsymbol{A}^{-1}Y = X_1 + \boldsymbol{A}^{-1}\boldsymbol{B}X_2 + \boldsymbol{A}^{-1}N$$

Thus the stated minimization problem can be simplified to

$$\min_{\hat{X}_1,\hat{X}_2} \left\| \boldsymbol{A}^{-1}Y - (\hat{X}_1 + \boldsymbol{A}^{-1}\boldsymbol{B}\hat{X}_2) \right\|_2$$

For extracting the data of the $i^{\text{th}}$ user, the best estimation of $X_1$ is as follows

1- If $i \le l$

$$\begin{cases} (\hat{X}_1)_j = sign\left((\boldsymbol{A}^{-1}Y - \boldsymbol{A}^{-1}\boldsymbol{B}\hat{X}_2)_j\right) & j = i \\ (\hat{X}_1)_j = softlim\left((\boldsymbol{A}^{-1}Y - \boldsymbol{A}^{-1}\boldsymbol{B}\hat{X}_2)_j\right) & j \ne i \end{cases}$$

In this case $\hat{X}_2$ takes all vectors in $\{-1,0,+1\}^{k-l}$.

2- If $i > l$

$$(\hat{X}_1)_i = softlim\left((\boldsymbol{A}^{-1}Y - \boldsymbol{A}^{-1}\boldsymbol{B}\hat{X}_2)_i\right)$$

In this case all entries of $\hat{X}_2$ belongs to $\{-1,0,+1\}$ except for the $i^{\text{th}}$ entry of $\hat{X}$ (this corresponds to the $(i-l)^{\text{th}}$ entry of $\hat{X}_2$) which only takes the values $\pm 1$.

Where $softlim(x)$ acts as a soft limiter and is defined by

$$softlim(x) = \begin{cases} -1 & x < -\frac{1}{2} \\ 0 & -\frac{1}{2} \le x \le +\frac{1}{2} \\ +1 & +\frac{1}{2} < x \end{cases}$$

When the input of $softlim(.)$ is a vector the function acts on each vector element. Thus, instead of looking between all likely estimates of $X$ (as stated earlier there are $2 \times 3^{k-1}$ such estimates) we need to only look between likely estimates of $X_2$ (there are either $2 \times 3^{k-1-l}$ or $3^{k-l}$such estimates). If $\boldsymbol{A}$ is a Hadamard matrix, it can easily be shown that the above algorithm is ML (because $\frac{1}{\sqrt{l}}\boldsymbol{A}$ is a unitary matrix and thus does not change the relative position of the points and also $softlim(X)$ is the nearest vector to $X$ with entries belonging to $\{-1,0,+1\}$).

**Example 2** Given a code matrix $\boldsymbol{E}_{64\times 88} = \boldsymbol{H}_4 \otimes \boldsymbol{C}_{16\times 22}$ where $\boldsymbol{H}_4$ is a $4 \times 4$ Hadamard matrix and $\boldsymbol{C}_{16\times 22}$ is the matrix defined in Table. 1. For the direct implementation of the ML decoder of $\boldsymbol{E}$, for each user, we need to look amongst $2 \times 3^{87}$ likely vectors and choose the one with minimum distance from the received vector. This shows that the direct ML decoder requires tremendous computational complexity and thus its implementation is not possible. By using the first major step, as stated above, this decoding problem reduces to 4 smaller decodings of $\boldsymbol{C}_{16\times 22}$. In this case we need to look amongst an overall number of $2 \times 3^{21}$ likely vectors for decoding the bit of each user. Furthermore, by using the second major step, we only have to look amongst at most $3^6$ likely vectors. This implies a tremendous reduction in complexity, making this decoder practically implementable. Notice that since in the definition of $\boldsymbol{E}$, $\boldsymbol{H}_4$ is a Hadamard matrix and also the first 16 columns of $\boldsymbol{C}_{16\times 22}$ is a Hadamard matrix, the overall decoder is still ML (according to the conditions discussed in the two major steps, under which using the two steps keeps the overall decoder ML).

## V. Simulation Results

To show the behavior of COWDA codes with respect to additive noise we simulated a CDMA system with 64 chips and 88 users in the presence of AWGN. The BER versus $E_b/N_0$ is depicted in Fig. 2. The curve for the simulation of the COWDA codes was obtained by using input vectors $X$ with equiprobable entries in the set $\{-1,0,+1\}$. Of course, this is a terrible case that does not occur in practice, because in actual systems when a user is active it stays active for a period of time. Thus, entries at different times are not independent and do not randomly alternate between $\pm 1$ and 0. Conclusively, in the simulated scenario at any time instant, the active users are completely unknown and are independent of the active users of the previous time instant. Therefore, the BER curve of COWDA is an upper bound on its performance in practical situations. To the extent of our knowledge no coding/decoding pair exists that does not need to know the active users for proper decoding (in the over-loaded case). Hence, there are no appropriate coding/decoding schemes for comparison. However, we have compared COWDA/ML decoder from the BER point of view with COW/ML decoder and BWBE/iterative decoder; obviously, this is not a fair comparison because in the latter two cases we have considered the case that the receiver knows that all users are active. That is, even if a user is not active the transmitter has to send a $+1$ or $-1$ as a fill code.

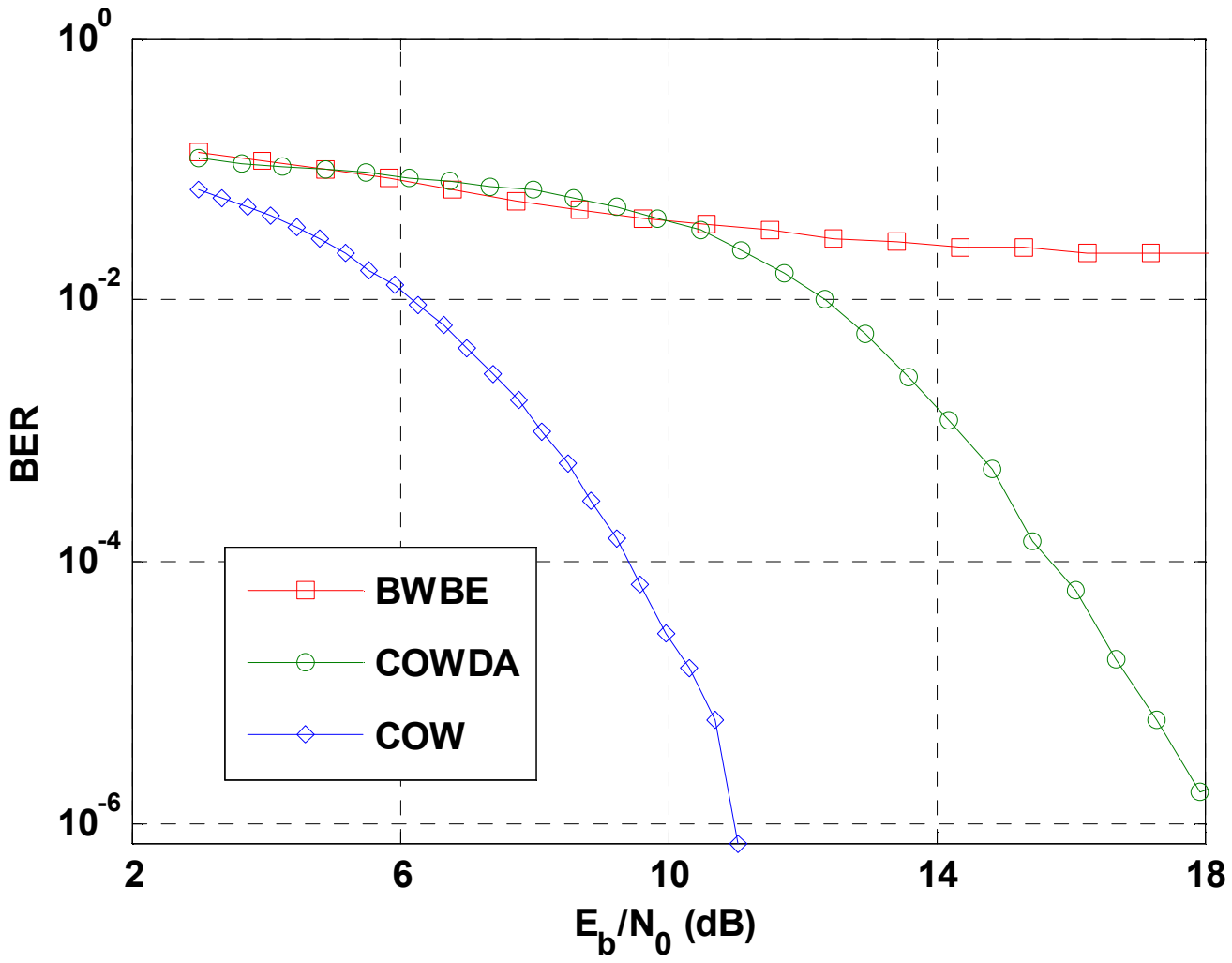

Fig. 2. Bit Error Rate (BER) versus $E_b/N_0$ for a CDMA system with 64 chips and 88 users.

As can be seen in Fig. 2, BWBE codes cannot reach error probabilities bellow a certain threshold even by greatly increasing the power. As obvious, in the curves the BER of COWDA tends to zero as $E_b/N_0$ increases. This means that we can have a CDMA system with as small as desired BER using COWDA codes. Despite the fact that COW codes seem to work better than COWDA codes (from the BER point of view), it should be taken notice that COW codes are not able to work in a system with unknown number of active users. This is the reason behind their ostensible better performance.

## VI. Conclusion

In this paper, we have shown that there exists a large class of errorless codes that are suitable for over-loaded synchronous CDMA which are also capable of detecting active users. We also proposed methods for large-size construction of these codes. In addition, we derived an upper-bound on the over-loading factor of these codes. We also presented an ML decoding scheme with acceptable computational complexity for a subclass of these codes. For example, we proposed a CDMA system with 64 chips that can handle 88 users, can achieve any desired BER, and can also detect the active users. Furthermore, the ML decoder of this code is computationally feasible and simulation results indicated that this scheme is robust against additive noise.

As described in section V, we studied the behavior of our coding/decoding scheme under the condition that input entries at different times are chosen from the set $\{-1,0,+1\}$ according to a uniform independent distribution. However, in practical situations such a terrible case does not occur at all. In future works, we intend to study the behavior of these codes (and to find the Maximum A Posterior decoder) under more practical patterns for activation/deactivation of the users which can result in much better performance in terms of BER.

## Appendix

We prove this theorem in 5 steps. Define $\boldsymbol{M} = [m_{ij}] = \boldsymbol{H}_2 \otimes \boldsymbol{C}$ and $\boldsymbol{D} = \boldsymbol{H}_2 \otimes \boldsymbol{M}$, and $\boldsymbol{\mathcal{L}} = \{\boldsymbol{D}X \mid X \in \{0, \pm1, \pm2\}^{4n}\}$, also $\boldsymbol{\mathcal{L}}' = \{\boldsymbol{M}X \mid X \in \{0, \pm1, \pm2\}^{2n}\}$.

***Step 1*** An interesting observation is that if $Z \in \{-1,+1\}^{4m}$ and $Z \notin \boldsymbol{\mathcal{L}}$ and $2Z \notin \boldsymbol{\mathcal{L}}$, then the augmented matrix $[\boldsymbol{D}|Z]$ is a COWDA matrix. The proof of this step is trivial.

***Step 2*** We would like to prove that if $\boldsymbol{\mathcal{B}} = Q + \{-1,+1\}^{4m}$, where $Q$ is an arbitrary $4m \times 1$ integer vector, then $|\boldsymbol{\mathcal{L}} \cap \boldsymbol{\mathcal{B}}| \leq 2^{2m+1}$. To show this, suppose that $Y \in \boldsymbol{\mathcal{L}} \cap \boldsymbol{\mathcal{B}}$. Then there exists a $\{0, \pm1, \pm2\}$-vector $X_{4n\times1} = [X_1{}^{\mathrm{T}} \quad X_2{}^{\mathrm{T}}]^{\mathrm{T}}$, where $X_1, X_2 \in \{0, \pm1, \pm2\}^{2n}$ and $Y = \boldsymbol{D}X$.

$Y = \boldsymbol{D}X = \begin{bmatrix} +\boldsymbol{M} & +\boldsymbol{M} \\ +\boldsymbol{M} & -\boldsymbol{M} \end{bmatrix} \begin{bmatrix} X_1 \\ X_2 \end{bmatrix} = \begin{bmatrix} \boldsymbol{M}X_1 + \boldsymbol{M}X_2 \\ \boldsymbol{M}X_1 - \boldsymbol{M}X_2 \end{bmatrix}$ and $Y_i = \boldsymbol{M}X_i$, for $i = 1, 2$; thus $Y = [(Y_1 + Y_2)^{\mathrm{T}} \quad (Y_1 - Y_2)^{\mathrm{T}}]^{\mathrm{T}}$.

Since there is a one-to-one correspondence between the set of vectors $[(Y_1 + Y_2)^{\mathrm{T}} \quad (Y_1 - Y_2)^{\mathrm{T}}]^{\mathrm{T}}$ and the set of vectors $[Y_1{}^{\mathrm{T}} \quad Y_2{}^{\mathrm{T}}]^{\mathrm{T}}$, the cardinality of the two sets are equal. Denote the $i^{\text{th}}$ entry of $Y_1$ by $(Y_1)_i$, thus we have

$$(Y_1)_i = \sum_{j=1}^{2n} m_{ij} (X_1)_j = \sum_{j=1}^{2n} (X_1)_j \pmod 2.$$

Hence the entries of $Y_1$ are either all odd or all even. Also this holds for $Y_2$. Since $Y \in \boldsymbol{\mathcal{B}}$, then for every $i$, $1 \leq i \leq 2m$, we have $\begin{cases} (Y_1)_i + (Y_2)_i = Q_i \pm 1 \\ (Y_1)_i - (Y_2)_i = Q_{2m+i} \pm 1 \end{cases}$.

By an easy calculation the solutions of the above equations are

$$\begin{cases} (Y_1)_i = \frac{Q_i + Q_{2m+i}}{2} \pm 1, (Y_2)_i = \frac{Q_i - Q_{2m+i}}{2} \\ (Y_1)_i = \frac{Q_i + Q_{2m+i}}{2}, (Y_2)_i = \frac{Q_i - Q_{2m+i}}{2} \pm 1 \end{cases}.$$

The above solutions are in two categories. Category 1 consists of the solutions which have 2 choices for $(Y_1)_i$ and only one choice for $(Y_2)_i$, while category 2 consists of solutions with a single choice for $(Y_1)_i$ and 2 choices for $(Y_2)_i$.

Now, for the determination of $|\boldsymbol{\mathcal{L}} \cap \boldsymbol{\mathcal{B}}|$, first assume that all entries of $Y_1$ are even and $l$ entries of $Y_1$ have two choices. Hence, the number of $[Y_1{}^{\mathrm{T}} \quad Y_2{}^{\mathrm{T}}]^{\mathrm{T}}$ vectors are $2^l 2^{2m-l} = 2^{2m}$, because the $l$ corresponding elements in $Y_2$ have only one choice and the other $2m - l$ elements in $Y_2$ have 2 choices. The same assertion holds when all entries of $Y_1$ are odd. Thus, $|\boldsymbol{\mathcal{L}} \cap \boldsymbol{\mathcal{B}}|$ has at most $2^{2m} + 2^{2m} = 2^{2m+1}$ elements.

***Step 3*** We would like to prove that if $\boldsymbol{\mathcal{P}} = T + \{0, \pm2\}^{2m}$, where $T$ is an arbitrary $2m \times 1$ integer vector, then $|\boldsymbol{\mathcal{L}}' \cap \boldsymbol{\mathcal{P}}| \leq 2^{2m+1}$. To show this, suppose that $Y \in \boldsymbol{\mathcal{L}}' \cap \boldsymbol{\mathcal{P}}$. Then there exists a $\{0, \pm1, \pm2\}$-vector $X_{2n\times1} = [X_1{}^{\mathrm{T}} \quad X_2{}^{\mathrm{T}}]^{\mathrm{T}}$, where $X_1, X_2 \in \{0, \pm1, \pm2\}^{n}$ and $Y = \boldsymbol{M}X$.

$Y = \boldsymbol{M}X = \begin{bmatrix} +\boldsymbol{C} & +\boldsymbol{C} \\ +\boldsymbol{C} & -\boldsymbol{C} \end{bmatrix} \begin{bmatrix} X_1 \\ X_2 \end{bmatrix} = \begin{bmatrix} \boldsymbol{C}X_1 + \boldsymbol{C}X_2 \\ \boldsymbol{C}X_1 - \boldsymbol{C}X_2 \end{bmatrix}$ and $Y_i = \boldsymbol{C}X_i$ for $i = 1, 2$; thus $Y = [(Y_1 + Y_2)^{\mathrm{T}} \quad (Y_1 - Y_2)^{\mathrm{T}}]^{\mathrm{T}}$ .

Since there is a one-to-one correspondence between the set of vectors $[(Y_1 + Y_2)^{\mathrm{T}} \quad (Y_1 - Y_2)^{\mathrm{T}}]^{\mathrm{T}}$ and the set of vectors $[Y_1{}^{\mathrm{T}} \quad Y_2{}^{\mathrm{T}}]^{\mathrm{T}}$, the cardinality of the two sets are equal. Denote the $i^{\text{th}}$ entry of $Y_1$ by $(Y_1)_i$, and the entry in the $i^{\text{th}}$ row and $j^{\text{th}}$ column of $\boldsymbol{C}$ by $c_{ij}$, thus we have

$$(Y_1)_i = \sum_{j=1}^{2n} c_{ij} (X_1)_j = \sum_{j=1}^{2n} (X_1)_j \pmod 2.$$

Hence the entries of $Y_1$ are either all odd or all even. Also this holds for $Y_2$. Since $Y \in \boldsymbol{\mathcal{P}}$, then for every $i$, $1 \leq i \leq m$, we have $\begin{cases} (Y_1)_i + (Y_2)_i = Q_i \pm a \\ (Y_1)_i - (Y_2)_i = Q_{m+i} \pm b \end{cases}$ where $a, b \in \{-2, 0, +2\}$.

By an easy calculation the solutions of the above equations are

$$\begin{cases} (Y_1)_i = \frac{Q_i + Q_{m+i}}{2} \pm 2, (Y_2)_i = \frac{Q_i - Q_{m+i}}{2} \\ (Y_1)_i = \frac{Q_i + Q_{m+i}}{2}, (Y_2)_i = \frac{Q_i - Q_{m+i}}{2} \pm 2 \end{cases}$$

or

$$(Y_1)_i = \frac{Q_i + Q_{m+i}}{2} \pm 1, (Y_2)_i = \frac{Q_i - Q_{m+i}}{2} \pm 1$$

Now, for the determination of $|\mathcal{L}' \cap \mathcal{P}|$, if all entries of $Y_1$ are odd then there exists at most four choices for the ordered pair $((Y_1)_i, (Y_2)_i)$. Thus for $Y_1$ and $Y_2$ we have at most $4^m$ choices. The same assertion holds when all entries of $Y_1$ are odd. Thus, $|\mathcal{L}' \cap \mathcal{P}|$ has at most $4^m + 4^m = 2^{2m+1}$ elements.

***Step 4*** We would like to prove that if $\mathcal{F} = Q + \{-2, +2\}^{4m}$, where $Q$ is an arbitrary $4m \times 1$ integer vector, then $|\mathcal{L} \cap \mathcal{F}| \leq 2^{3m} + 2^{m-1}$. To show this, suppose that $Y \in \mathcal{L} \cap \mathcal{F}$. Then there exists a $\{0, \pm1, \pm2\}$-vector $X_{4n\times1} = [X_1{}^{\mathrm{T}} \quad X_2{}^{\mathrm{T}}]^{\mathrm{T}}$, where $X_1, X_2 \in \{0, \pm1, \pm2\}^{2n}$ and $Y = \boldsymbol{D}X$.

$$Y = \boldsymbol{D}X = \begin{bmatrix} +\boldsymbol{M} & +\boldsymbol{M} \\ +\boldsymbol{M} & -\boldsymbol{M} \end{bmatrix} \begin{bmatrix} X_1 \\ X_2 \end{bmatrix} = \begin{bmatrix} \boldsymbol{M}X_1 + \boldsymbol{M}X_2 \\ \boldsymbol{M}X_1 - \boldsymbol{M}X_2 \end{bmatrix} \text{ and } Y_i = \boldsymbol{M}X_i$$

for $i = 1, 2$; thus $Y = [(Y_1 + Y_2)^{\mathrm{T}} \quad (Y_1 - Y_2)^{\mathrm{T}}]^{\mathrm{T}}$.

Since there is a one-to-one correspondence between the set of vectors $[(Y_1 + Y_2)^{\mathrm{T}} \quad (Y_1 - Y_2)^{\mathrm{T}}]^{\mathrm{T}}$ and the set of vectors $[Y_1{}^{\mathrm{T}} \quad Y_2{}^{\mathrm{T}}]^{\mathrm{T}}$, the cardinality of the two sets are equal. Denote the $i^{\text{th}}$ entry of $Y_1$ by $(Y_1)_i$, if $Y \in \mathcal{F}$, then for every $i$, $1 \leq i \leq 2m$, we have

$$\begin{cases} (Y_1)_i + (Y_2)_i = Q_i \pm 2 \\ (Y_1)_i - (Y_2)_i = Q_{2m+i} \pm 2 \end{cases}.$$

By an easy calculation the solutions of the above equations are

$$\begin{cases} (Y_1)_i = \frac{Q_i + Q_{2m+i}}{2} \pm 2, (Y_2)_i = \frac{Q_i - Q_{2m+i}}{2} \\ (Y_1)_i = \frac{Q_i + Q_{2m+i}}{2}, (Y_2)_i = \frac{Q_i - Q_{2m+i}}{2} \pm 2 \end{cases}.$$

Define the $2m \times 1$column vector $T$ and $R$ where the $i^{\text{th}}$ entry is given by $T_i = \frac{Q_i + Q_{2m+i}}{2}$ and $R_i = \frac{Q_i - Q_{2m+i}}{2}$.

Define $\mathcal{H} = \{V | V \in \{-2, 0, +2\}^{2m}, T + V \in \mathcal{L}'\}$. For every $V$ in $\mathcal{H}$, there exists at most $2^{|nonezero\ entries\ of\ V|}$ vectors $W$ in $\{-2, 0, +2\}^{2m}$ such that $Y_1 = T + V$ and $Y_2 = R + W$. Thus,

$$|\mathcal{L} \cap \mathcal{F}| \leq \sum_{V \in \mathcal{H}} 2^{|nonezero\ entries\ of\ V|}$$

The summation of the right side of the above inequality can be rewritten as

$$\sum_{k=0}^{2m} 2^k (members\ of\ \mathcal{H}\ with\ k\ zero\ entries) \leq 2^m + (|\mathcal{H}| - 1)2^{m-1}$$

According to step 3 the cardinality of $\mathcal{H}$ is at most $2^{2m+1}$. Hence, $|\mathcal{L} \cap \mathcal{F}| \leq 2^{3m} + 2^{m-1}$.

***Step 5*** Now, suppose that we add $k$ columns to $\boldsymbol{D}$, $k < \lceil (m-1) \log_5 2 \rceil$, and the resultant matrix, $\boldsymbol{E}$, is a COWDA matrix. We wish to prove that one can add another column to $\boldsymbol{E}$ to obtain a COWDA matrix with $4n + k + 1$ columns. Assume that $\boldsymbol{E} = [\boldsymbol{D}|\boldsymbol{F}]$, where $\boldsymbol{F} = [W_1|\cdots|W_k]$, and $W_i$ is a $4m \times 1$ vector, for $i = 1, \ldots, k$. Let $X \in \{0, \pm1, \pm2\}^{4n+k}$, $X = [X_1{}^{\mathrm{T}} \quad X_2{}^{\mathrm{T}}]^{\mathrm{T}}$, where $X_1$ is a $4n \times 1$ vector and $X_2$ is a $k \times 1$ vector. Hence, $\boldsymbol{D}X_1 = \boldsymbol{E}X - \boldsymbol{F}X_2$. By steps 2 and 4, and the fact that $X_2$ has $5^k$ different possibilities, we have

$$|(\boldsymbol{\mathcal{V}} \cap \{-1, +1\}^{4m}) \cup (\boldsymbol{\mathcal{V}} \cap \{-2, +2\}^{4m})|$$
$$= \sum_{X_2} |\{\boldsymbol{D} \cdot \{0, \pm1, \pm2\}^{4n}\} \cap (\{-\boldsymbol{F}X_2 + \{-1, +1\}^{4m}\} \cup \{-\boldsymbol{F}X_2 + \{-2, +2\}^{4m}\})| \leq 5^k (2^{3m} + 2^{m-1} + 2^{2m+1})$$
$$\leq 5^k 2^{3m+1}$$

where $\boldsymbol{\mathcal{V}} = \{\boldsymbol{E}X | X \in \{0, \pm1, \pm2\}^{4n+k}\}$.

Now, if $5^k 2^{3m+1} < |\{-1, +1\}^{4m}| = 2^{4m}$, then we can add another column to matrix $\boldsymbol{D}$ by applying step 1. Thus, we can add at least $\lceil (m-1) \log_5 2 \rceil$ vectors to $\boldsymbol{D}$ and obtain a bigger COWDA matrix. ■